\pdfoutput=1
\documentclass[12pt]{article}
\usepackage[english]{babel}
\usepackage{color}
\usepackage{graphicx}
\usepackage{amsmath}
\usepackage{amsthm}
\usepackage{amssymb}
\usepackage[hyperindex,plainpages=false]{hyperref}
\usepackage[left=1.1in,right=1.1in,bottom=1in,top=1in,a4paper]{geometry}

\title{Four Symmetries of the KdV equation}
\author{Alexander G. Rasin  \\
Department of Computer Science and Mathematics,\\ 
Ariel University, Ariel 40700, Israel \\
{E-mail: rasin@ariel.ac.il}\and  Jeremy Schiff (Corresponding author) \\
Department of Mathematics,\\
Bar-Ilan University, Ramat Gan, 52900, Israel \\
{E-mail: schiff@math.biu.ac.il}}

\begin{document}
\maketitle
\begin{abstract}
   We revisit the symmetry structure of integrable PDEs, looking at the specific example of the KdV equation.
  We identify 4 nonlocal symmetries of KdV depending on a parameter,  which we call generating symmetries.
  We explain that since these are nonlocal  symmetries, their commutator algebra is not uniquely determined, and we present three
  possibilities for the algebra. In the first version, 3 of the 4 symmetries commute; this shows that it is possible to
  add further (nonlocal) commuting flows to the standard KdV hierarchy. The second version of the commutator algebra
  is consistent with Laurent expansions of the symmetries, giving rise to an infinite dimensional algebra of
  hidden symmetries of KdV. The third version is consistent with asymptotic expansions for large values of the
  parameter, giving rise to the standard commuting symmetries of KdV, the infinite hierarchy of
  ``additional symmetries'', and their traditionally accepted commutator algebra (though this also
  suffers from some ambiguity as the additional symmetries are nonlocal). 
  We explain how the 3 symmetries that commute in the first version of the algebra can all be regarded as
  infinitesimal double B\"acklund transformations. The 4 generating symmetries 
  incorporate all known symmetries of the KdV equation, but also exhibit some remarkable novel structure, 
  arising from their nonlocality. We believe this structure to be shared by other integrable PDEs.
\end{abstract}

\section{Introduction} 

    Symmetry methods \cite{Ov, Ol0, BK, BCA}, dating back to the work of Lie and Noether, remain the main
    tool for finding explicit solutions of differential equations, and identifying and using 
    the full symmetry structure of a given differential equation (or system
    of differential equations), is essential for developing both appropriate mathematical tools
    and insight into whatever the equation is being used to describe. 
    It has become appreciated how important it is to take symmetry into 
    account when choosing numerical methods for differential equations \cite{HLW}, and even simple symmetries 
    such as  translations and scalings,  play a central role in various areas, such as 
    discussion of the stability of solitons \cite{Tao} and computing the quantum effects due to instantons in gauge
    theory \cite{tHooft}.

    A significant number of nonlinear partial differential equations (PDEs)  that arise in applications, ranging from particle physics
    and gravity to fluid mechanics and optics, exhibit a set of special  properties that earn them the sobriquet ``integrable''.
    One of these properties is that they exhibit an infinite hierarchy of symmetries. As pointed out in \cite{Ol1},
    this was evident for the Korteweg-de Vries (KdV) equation already from the results of \cite{Gard}. 
    But it was first demonstrated explicitly for the sine-Gordon equation
    using a generating function technique by Kumei \cite{Kumei}, and then for the KdV, Burgers, modified KdV and sine-Gordon equations 
    using the technique of recursion operators by Olver \cite{Ol1}.  And in the 45 years since these pioneering papers, hierarchies
    of symmetries have been identified for countless other integrable systems.

For many well-studied systems it seems  that there is more than a single hierarchy of symmetries. 
For the case of the  KdV equation, there are scaling and Galilean symmetries, and application of the recursion operator to  these leads to a second infinite hierarchy,
known as the additional symmetries \cite{IS,OS,Khor}.  Unlike the standard symmetries,
the additional symmetries do not commute among themselves or with the standard symmetries, and they are nonlocal
(the meaning of this will be explained shortly). As well as the additional symmetries, various other nonlocal hidden symmetries
have been identified  \cite{Guthrie,GH,Lou,OC2,LHC}. 

The research reported in this paper is the outgrowth of our attempt to answer a simple question.  In \cite{RSG},
following in the footsteps of Kumei \cite{Kumei},  we 
described a generating function for the standard commuting symmetries of KdV; that is, we found a symmetry depending on a
parameter, which, when expanded in an asymptotic series for large values of the parameter, gave the full standard hierarchy
of symmetries.  The existence of such a ``generating symmetry'' for KdV was known before, and it is known as ``the square eigenfunction
symmetry'' or as ``the resolvent''\cite{OC2,Dic}.
In \cite{RSG} we showed how to interpret this as an infinitesimal double  B\"acklund transformations of KdV,
giving substantial simplifications, including a simple proof of commutativity. The original question we set out to answer now
was {\em is it possible to write a generating function for the additional symmetries of KdV?} For the KP hierarchy
such a generating function is known \cite{Dickeypap}, though it is far from obvious how to reduce it to the case of KdV.  
After significant  effort, we succeeded to find a succinct form of the relevant 
generating function for KdV. But we also found more: In section 2 of this paper we show the existence of 4 distinct
symmetries of KdV depending on a parameter. 

These symmetries are all nonlocal. For local symmetries, the characteristics can be written as explicit functions of  the independent variables,
the dependent variables, and their derivatives. For nonlocal symmetries there is a more general dependence, for example on integrals of
the dependent variables, or potential functions from which solutions of the relevant PDEs can be found.  Confirming that a given
expression defines a nonlocal symmetry is no harder than doing so for a local symmetry. But to compute the commutator algebra, for
example, is problematic, typically because of the lack of single-valuedness of the action of nonlocal symmetries associated with undefined
constants of integration. Over the years this subject has been extensively investigated, amongst others by the groups of Krasilshchik and
Vinogradov \cite{KV1,KV2}  and Bluman and Cheviakov \cite{bc1,bc2,bc3}. In our case, the characteristics of the 4 generating
symmetries are expressed in terms of a single potential
depending on the parameter. To compute the commutator algebra, we need to fix the actions of the 4 symmetries on this potential.
These are not fully determined {\it ab initio}, and indeed there are different ways to do this, giving rise to distinct commutator algebras. 
(The fact that this happens, in some generality, for nonlocal symmetries, is not widely appreciated; though similar issues, arising in 
checking the Jacobi identity for nonlocal symmetries, were addressed in \cite{OSW,Ol2}.)  
In section 3  we look at three possible  choices of these actions and present the associated algebras.
For the first choice, 3 of the 4 symmetries commute. The implication of this is that {\em there are further (nonlocal) flows that can
be added to the KdV hierarchy}. We illustrate this unexpected result very explicitly.
The second and third choices avoid a certain singularity, which, we speculate, is important for there to be consistent series
representations of the symmetries.

In section 4 of the paper we look at series expansions of the 4 generating symmetries  for large and small values of the parameter, and also in Laurent series.
Asymptotic expansions of 2 of the symmetries for large values of the parameter  give the standard hierarchy and the
hierarchy of additional symmetries.  
Expansions in power series for small values of the parameter give hierarchies of hidden symmetries, as 
identified by Guthrie and Hickman in \cite{GH,Guthrie} and by Lou in \cite{Lou}.
The components in these expansions are related by a recursion relation,  
arising from certain identities satisfied by the 4 symmetries. 
It seems that the different expansions are all only consistent with one 
specific version of the commutator algebra.
We obtain  commutators for the standard hierarchy  and additional symmetries from the
third version of the algebra from section 3, and 
a commutator algebra for the hidden symmetries from the second version
of the algebra from section 3. 

In section 5 of the paper, we explain that the 3 of the 4 symmetries that commute (with the appropriate choice for their extended actions) 
are actually all infinitesimal double B\"acklund transformations.  Section 6 contains some concluding comments  and questions for further study.

The results of this paper are all specific for the KdV equation, the archetype of integrable $1+1$ dimensional PDEs. 
However, as we will explain in  section 2, the form of the symmetries suggests a simple way to search for generating symmetries
for other PDEs, and we have already had some success in this   direction \cite{SashaPreprint}.  
For the case of KdV it seems that the 4 generating symmetries encode 
everything known about symmetries of the KdV equation.  Thus our work raises the prospect that in some generality it may be possible to
describe the symmetry structure of an integrable $1+1$ dimensional PDE using a finite set of generating symmetries. This
would be a significant advance.

\section{The 4 Symmetries} 

We work with the potential KdV equation (pKdV)
\begin{equation}
  u_t - \textstyle{\frac32} u_x^2 - \textstyle{\frac14} u_{xxx} = 0 \ . \label{pKdV} 
\end{equation}  
The characteristic of a symmetry is any quantity $\eta$ that satisfies the linearized equation
$$ \eta_t  - 3 u_x \eta_x  - \textstyle{\frac14} \eta_{xxx} = 0   $$
whenever $u$ satisfies pKdV.

\medskip
\noindent{\bf Theorem 1:}
Let $z_\alpha$ be a solution of the system  
\begin{eqnarray}
  4u_x &=& \frac{z_{\alpha,xxx}}{z_{\alpha,x}} - \frac32 \frac{z_{\alpha,xx}^2}{z_{\alpha,x}^2} + 2 \alpha - \frac12 z_{\alpha,x}^2    \ , \label{zadef}\\
   z_{\alpha,t}  &=&  (\alpha+u_x) z_{\alpha,x} \ ,  \label{zat}
\end{eqnarray}
where $\alpha$ is a parameter. Then the following are characteristics of symmetries of pKdV: 
\begin{eqnarray}
  Q(\alpha) &=& \frac{1}{z_{\alpha,x}} \label{Qdef} \ ,  \\   
  R(\alpha) &=&   \frac{z_{\alpha,\alpha}}{z_{\alpha,x}} - \frac32 t  \ ,   \label{Rdef} \\ 
  S(\alpha) &=& \frac{e^{z_\alpha}}{z_{\alpha,x}}\ ,  \label{Sdef} \\ 
  T(\alpha) &=& \frac{e^{-z_\alpha}}{z_{\alpha,x}} \ .  \label{Tdef}  
\end{eqnarray}
($z_{\alpha,\alpha}$ denotes the derivative of $z_\alpha$ with respect to $\alpha$.)  

\medskip
\noindent{\bf Proof:} A direct computation.  $\bullet$ 

\medskip 
\noindent{\bf Notes:} 
\newline 1.  The consistency condition for the system (\ref{zadef})-(\ref{zat}) is the KdV equation (i.e. the $x$ derivative of equation (\ref{pKdV})).
         In fact the system (\ref{zadef})-(\ref{zat}) is just a rather unobvious way to write the Lax pair of the KdV equation (see Note 3 below),
           or, equivalently, the system (\ref{BT10})-(\ref{BT11}) that determines a B\"acklund transformation of the KdV equation (see equation (\ref{vtoz}) 
           that shows how a solution of (\ref{zadef})-(\ref{zat}) gives rise to two solutions of (\ref{BT10})-(\ref{BT11})). It is possible to write the
           characteristics $Q,R,S,T$, and all the other results of this paper
           in terms of solutions of the Lax pair or the system defining B\"acklund transformations, but the manipulations are somewhat easier using $z_\alpha$.
\newline 2.  The system (\ref{zadef})-(\ref{zat}) is clearly invariant under the discrete symmetry  $z_\alpha \leftrightarrow -z_\alpha$ which switches
$S$ and $T$.  In greater generality, if we write $z_\alpha = \log w_\alpha$, the system is invariant under the action of M\"obius transformations on $w_\alpha$.
Once a single solution $z_\alpha$ of the system (\ref{zadef})-(\ref{zat}) is known, the general solution can be written in the form  
    $$  \log\left(  \frac{  C_1 e^{z_\alpha} + C_2}{  C_3 e^{z_\alpha} + C_4 }  \right) \ ,    $$
where $C_1,C_2,C_3,C_4$ are constants.
\newline 3.  Writing $\psi^{(1)}_\alpha = \sqrt{S(\alpha)}$  and  $\psi^{(2)}_\alpha = \sqrt{T(\alpha)}$ it is straightforward to verify that
$\psi^{(1)}_\alpha, \psi^{(2)}_\alpha$ are both solutions of the KdV Lax pair 
\begin{eqnarray*}
  \psi_{\alpha,xx} &=& (\alpha-2u_x)  \psi_\alpha \ , \\
  \psi_{\alpha,t}  &=& (\alpha+u_x) \psi_{\alpha,x}- \frac12 u_{xx}\psi_\alpha  \ .
\end{eqnarray*}
Thus  the three symmetries $Q(\alpha),S(\alpha),T(\alpha)$ are the three ``square eigenfunction'' symmetries $\psi^{(1)}_\alpha \psi^{(2)}_\alpha ,  \left(  \psi^{(1)}_\alpha  \right)^2
,  \left(  \psi^{(2)}_\alpha  \right)^2 $. These are not new \cite{OC2,Dic}. As far as we know, though, the symmetry $R(\alpha)$ is new, as is the expression of all the
symmetries in terms of $z_\alpha$, which  facilitates commutator computations, as will be shown in the next section.   
\newline 4. 
  We found the formula for $R(\alpha)$ by a long process of trial and error.
  $R(\alpha)$ can also be expressed in terms of $\psi^{(1)}_\alpha$ and  $\psi^{(2)}_\alpha$ and their $\alpha$ derivatives,  
  but the expression in terms of the single function $z_\alpha$ is simpler for commutator computations.
  The expression in terms of solutions of the Lax pair suggests a method to search for an analogous symmetry for other integrable equations
  \cite{SashaPreprint}.
\section{Commutator Computations} 

Suppose that $\delta u$ is the variation of $u$ under a certain symmetry
(i.e. $\delta u$ is the characteristic of the symmetry).
Then for consistency with (\ref{zadef})-(\ref{zat}) the  variation $\delta z_{\alpha}$ of $z_\alpha$  must satisfy the linear system
\begin{eqnarray}
  4{\delta u}_x &=& \frac{{\delta z}_{\alpha,xxx}}{z_{\alpha,x}} -  \frac{z_{\alpha,xxx}{\delta z}_{\alpha,x} }{z_{\alpha,x}^2} 
  - 3 \frac{z_{\alpha,xx} {\delta z}_{\alpha,xx} }{z_{\alpha,x}^2} +  3 \frac{z_{\alpha,xx}^2   {\delta z}_{\alpha,x}  }{z_{\alpha,x}^3} 
  -  z_{\alpha,x} {\delta z}_{\alpha,x}   \ , \label{zsym1} \\
   {\delta z}_{\alpha,t}  &=&  {\delta u}_x  z_{\alpha,x}  +  (\alpha+u_x) {\delta z}_{\alpha,x}  \ .   \label{zsym2}  
\end{eqnarray}
As explained in the introduction, this does not fully determine $\delta z_\alpha$ for a given  $\eta$.  However, it does immediately allow to check
consistency of a proposed form of $\delta z_\alpha$. 

\medskip 
\noindent{\bf Notation:}  We denote by $\delta_\eta u$ and $\delta_\eta z_\alpha$ the variations of $u$ and $z_\alpha$ respectively under
the symmetry with characteristic $\eta$.  Thus $\delta_\eta z_\alpha$ should be a solution of the system (\ref{zsym1})-(\ref{zsym2}) with \
${\delta u}=\eta$.  

\medskip 
\noindent{\bf Theorem 2:}  The following formulae give consistent actions of the symmetries $Q(\beta)$, $R(\beta),S(\beta),T(\beta)$ on $z_\alpha$: 
\begin{eqnarray}
  \delta_{Q(\beta)} z_\alpha  &=&  \frac{z_{\alpha,x}}{(\beta-\alpha)z_{\beta,x}}\ , \label{deltaQ}  \\    
  \delta_{R(\beta)} z_\alpha  &=& \frac{z_{\alpha,x} z_{\beta,\beta} - z_{\beta,x} z_{\alpha,\alpha} }{(\beta-\alpha)z_{\beta,x}}\ ,  \label{deltaR}  \\
  \delta_{S(\beta)} z_\alpha  &=& \frac{z_{\alpha,x}e^{z_\beta}}{(\beta-\alpha)z_{\beta,x}}\ ,  \label{deltaS} \\
  \delta_{T(\beta)} z_\alpha  &=& \frac{z_{\alpha,x}e^{-z_\beta}}{(\beta-\alpha)z_{\beta,x}}\ .   \label{deltaT}
\end{eqnarray}

\medskip
\noindent{\bf Proof:} A direct computation.
The relevant formulae for  $\delta_{Q(\beta)}u$ etc. are given by equations 
(\ref{Qdef})-(\ref{Tdef}) with $\alpha$ replaced by $\beta$.  $\bullet$ 

\medskip 
\noindent{\bf Notes:}
\newline 1. 
The simplicity of these formulae is remarkable. Once again, we will not present here any derivation, as once
the formulae are known they can be directly verified, and the derivation we currently have is not systematic. 
\newline 2.   These formulae for $ \delta_{Q(\beta)} z_\alpha, \delta_{S(\beta)} z_\alpha, \delta_{T(\beta)} z_\alpha $ 
diverge for $\beta=\alpha$.  Thus the commutator computations we present below are not valid for $\alpha=\beta$. 

\medskip\noindent
Once the variations of $z_\alpha$ are known it is straightforward to compute the commutators of the symmetries:  

\medskip 
\noindent{\bf Theorem 3:} Using formulae (\ref{deltaQ})-(\ref{deltaT}) we have, for $\alpha\not=\beta$:   
\begin{eqnarray}
&&  [Q(\beta),Q(\alpha)] =  [S(\beta),S(\alpha)] =   [T(\beta),T(\alpha)]  =   0 \ , \\ 
&&  [Q(\beta),S(\alpha)] =  [Q(\beta),T(\alpha)] =   [T(\beta),S(\alpha)]  =   0 \ , \\
  &&  [R(\beta),Q(\alpha)] = \frac{\partial}{\partial\alpha}\left( \frac{Q(\alpha)}{\alpha-\beta} \right) \ , \label{RQ1}\\
  &&  [R(\beta),S(\alpha)] =   \frac{\partial}{\partial\alpha}\left( \frac{S(\alpha)}{\alpha-\beta} \right)  \ , \\
  &&  [R(\beta),T(\alpha)]  =  \frac{\partial}{\partial\alpha}\left( \frac{T(\alpha)}{\alpha-\beta} \right)    \ ,  \\ 
&&  [R(\beta),R(\alpha)] =    \frac{2(R(\beta)-R(\alpha))}{(\alpha-\beta)^2}  + \frac{R'(\beta)+R'(\alpha)}{\alpha-\beta}  \ . 
\end{eqnarray}
Here we abuse notation by using the same symbol for a symmetry and its characteristic, and $R'(\alpha)$ denotes
the derivative of $R(\alpha)$ with respect to $\alpha$.   

\medskip
\noindent{\bf Proof:} By definition the characteristic of the commutator of, say, $Q(\beta)$ and $R(\alpha)$ is
$$         [Q(\beta), R(\alpha)] = \delta_{Q(\beta)} R(\alpha) -  \delta_{R(\alpha)} Q(\beta)\ .  $$
From  theorem 2 we know the action of  $Q(\beta),R(\beta),S(\beta),T(\beta)$ on $z_\alpha$, and 
$Q(\alpha),R(\alpha),S(\alpha),T(\alpha)$ are determined by $z_\alpha$, and its $x$ and $\alpha$ derivatives. Thus
it is straightforward to compute the action of $Q(\beta),R(\beta),S(\beta),T(\beta)$ also on 
$Q(\alpha),R(\alpha),S(\alpha),T(\alpha)$. For example
$$ \delta_{Q(\beta)} R(\alpha) =
\delta_{Q(\beta)} \left( \frac{z_{\alpha,\alpha}}{z_{\alpha,x}} - \frac32 t \right)  
=  \frac{(\delta_{Q(\beta)}z_\alpha)_\alpha} {z_{\alpha,x}} -  \frac{z_{\alpha,\alpha} (\delta_{Q(\beta)}z_\alpha)_x  }{z_{\alpha,x}^2}\ . 
$$
Verification of the commutators is thus reduced to a direct computation. $\bullet$

\medskip\noindent
Theorem 3 has a remarkable consequence.
From the fact that  $Q,S,T$ all commute it follows that {\em there are further (nonlocal) flows that can
  be added to the KdV hierarchy}.  We state a simple case of  this as a corollary that can checked directly without any reference to
the  other results in this paper: 

\medskip
\noindent{\bf Corollary:} The following flows are consistent, for a single function  $u(x,t,t_+,t_-)$:
\begin{eqnarray}
  \frac{\partial u}{\partial t} &=&  \frac14 u_{xxx} + \frac32  u_x^2  \ ,  \label{ut} \\
  \frac{\partial u}{\partial t_+} &=&  \frac{e^{z_\alpha}}{z_{\alpha,x}}  \ ,  \label{utp} \\
  \frac{\partial u}{\partial t_-} &=&  \frac{e^{-z_\beta}}{z_{\beta,x}}   \ . \label{utm} 
\end{eqnarray}  
Here  $z_\alpha(x,t,t_+,t_-)$ is a solution of (\ref{zadef})-(\ref{zat}),   $z_\beta(x,t,t_+,t_-)$ is a solution of the same system with the
parameter $\alpha$ replaced by a different parameter $\beta$,  and
\begin{eqnarray}
  \frac{\partial z_\alpha}{\partial t_-} &=& \frac1{\beta-\alpha}  \frac{z_{\alpha,x}}{z_{\beta,x}} e^{-z_\beta}\ ,  \label{zatm} \\ 
  \frac{\partial z_\beta}{\partial t_+} &=&  \frac1{\alpha-\beta}  \frac{z_{\beta,x}}{z_{\alpha,x}} e^{z_\alpha}\ .   \label{zbtp} 
\end{eqnarray}
(The latter equations are consistent with the $t$-flows of $z_\alpha,z_\beta$ given by  (\ref{zat}) and (\ref{zat}) with
$\alpha$ replaced by $\beta$.) 

\medskip\noindent{\bf Notes:}
\newline 1. 
Here the $t_+$ flow corresponds to the action of $S(\alpha)$ and the $t_-$ flow corresponds to the action of $T(\beta)$. A further
flow could be included corresponding to the action of $Q(\gamma)$. Since, as was shown in \cite{RSG}, $Q$ acts as a generating function
for the entire KdV hierarchy, the $t_+$ and $t_-$ flows also commute with the entire KdV hierarchy.   
\newline 2.
We note that to check the consistency of the system (\ref{ut})-(\ref{utp})-(\ref{utm}) there is no need 
for explicit formulae for
$$    \frac{\partial z_\alpha}{\partial t_+} \qquad {\rm and} \qquad   \frac{\partial z_\beta}{\partial t_-}  \ .    $$ 
This corresponds to the fact that we have no formulae for the action of $S(\alpha)$  or $T(\alpha)$ on $z_\alpha$, only for 
the action of $S(\beta)$  or $T(\beta)$ on $z_\alpha$ for $\beta\not=\alpha$.  

\medskip\noindent
Returning to our main subject,  as we have explained, equations (\ref{deltaQ})-(\ref{deltaT}) do not give the most
general possibility for the actions of the symmetries on $z_\alpha$.  It is straightforward to check that the general solution of the
homogeneous version of (\ref{zadef})-(\ref{zat})  (i.e. the system obtained from this by setting $\delta u=0$) is a linear
combination of  $1, e^{z_\alpha}, e^{-z_\alpha}$. Thus is is possible to add any linear combination of these terms to 
each of the variations in (\ref{deltaQ})-(\ref{deltaT}). The coefficients of these linear combinations can be functions
of the parameters $\alpha$ and $\beta$.  We look at two particular cases.  

\medskip 
\noindent{\bf Theorem 3$'$:} The following formulae also give consistent actions of the symmetries $Q(\beta)$, $R(\beta)$,  $S(\beta),T(\beta)$ on $z_\alpha$:
\begin{eqnarray}
  \delta_{Q(\beta)} z_\alpha  &=&  \frac{z_{\alpha,x}- z_{\beta,x}}{(\beta-\alpha)z_{\beta,x}}\ ,  \label{deltaQ2} \\    
  \delta_{R(\beta)} z_\alpha  &=& \frac{z_{\alpha,x} z_{\beta,\beta} - z_{\beta,x} z_{\alpha,\alpha} }{(\beta-\alpha)z_{\beta,x}}\ ,  \label{deltaR2} \\
  \delta_{S(\beta)} z_\alpha  &=& \frac{z_{\alpha,x}e^{z_\beta} -  z_{\beta,x}e^{z_\alpha}  }{(\beta-\alpha)z_{\beta,x}} \ ,   \label{deltaS2} \\    
  \delta_{T(\beta)} z_\alpha  &=& \frac{z_{\alpha,x}e^{-z_\beta}-  z_{\beta,x}e^{-z_\alpha}  }{(\beta-\alpha)z_{\beta,x}}\ .  \label{deltaT2}   
\end{eqnarray}
These give rise to the following commutator algebra: 
\begin{eqnarray}
&&  [Q(\beta),Q(\alpha)] =  [S(\beta),S(\alpha)] =   [T(\beta),T(\alpha)]  =   0 \ , \\ 
&&  [Q(\beta),S(\alpha)] =  \frac{ S(\alpha)- S(\beta)  }{\alpha-\beta}   \ ,   \label{QS2} \\
  &&  [Q(\beta),T(\alpha)] =    - \frac{ T(\alpha)-  T(\beta)  }{\alpha-\beta}   \ , \\
&&  [T(\beta),S(\alpha)]  =   \frac{2( Q(\alpha)-  Q(\beta))  }{\alpha-\beta}   \ ,     \\
  &&  [R(\beta),Q(\alpha)] = \frac{\partial}{\partial\alpha}\left( \frac{Q(\alpha)}{\alpha-\beta} \right) +  \frac{Q(\beta)}{(\alpha-\beta)^2}  \ ,  \label{RQ2}\\
  &&  [R(\beta),S(\alpha)] =   \frac{\partial}{\partial\alpha}\left( \frac{S(\alpha)}{\alpha-\beta} \right) +  \frac{S(\beta)}{(\alpha-\beta)^2}  \ , \\
  &&  [R(\beta),T(\alpha)]  =  \frac{\partial}{\partial\alpha}\left( \frac{T(\alpha)}{\alpha-\beta} \right) +  \frac{T(\beta)}{(\alpha-\beta)^2}   \ ,  \\ 
&&  [R(\beta),R(\alpha)] =    \frac{2(R(\beta)-R(\alpha))}{(\alpha-\beta)^2}  + \frac{R'(\beta)+R'(\alpha)}{\alpha-\beta} \label{RR} \ . 
\end{eqnarray}

\medskip \noindent 
Note that for the choice of the arbitrary constants in $\delta_Q,\delta_S,\delta_T$ in theorem 3$'$ the variations 
can also be defined for $\beta=\alpha$ (using l'H\^opital's rule).
However this is not the only choice for which this is true, as the constants can be taken to depend on the parameters
$\alpha,\beta$.  We illustrate using just the  $Q,R$ subalgebra (which will be sufficient for our applications later). 

\medskip 
\noindent{\bf Theorem 3$''$:} The following formulae also give consistent actions of the symmetries $Q(\beta),R(\beta)$   on $z_\alpha$:
\begin{eqnarray}
  \delta_{Q(\beta)} z_\alpha  &=&  \frac{z_{\alpha,x}- \frac{\beta^{1/2}}{\alpha^{1/2}}   z_{\beta,x}}{(\beta-\alpha)z_{\beta,x}}\ ,  \label{deltaQ3} \\    
  \delta_{R(\beta)} z_\alpha  &=& \frac{z_{\alpha,x} z_{\beta,\beta} - z_{\beta,x} z_{\alpha,\alpha} }{(\beta-\alpha)z_{\beta,x}}\ .  \label{deltaR3} 
\end{eqnarray}
These give rise to the following commutator algebra: 
\begin{eqnarray}
&&  [Q(\beta),Q(\alpha)]  =   0 \ , \\ 
  &&  [R(\beta),Q(\alpha)] = \frac{\partial}{\partial\alpha}\left( \frac{Q(\alpha)}{\alpha-\beta} \right) +
     \frac{(3\beta-\alpha)\alpha^{1/2}}{2\beta^{3/2}(\beta-\alpha)^2}  Q(\beta)      \ ,  \label{RQ3}\\
&&  [R(\beta),R(\alpha)] =    \frac{2(R(\beta)-R(\alpha))}{(\alpha-\beta)^2}  + \frac{R'(\beta)+R'(\alpha)}{\alpha-\beta} \label{RR3} \ . 
\end{eqnarray}

\section{Expansions in components} 

\subsection{Expansions for large values of $|\alpha|$} 

The fact that $Q(\alpha)$ and $R(\alpha)$ are generating functions of, respectively,  the standard commuting symmetries and the additional
symmetries of the KdV equation can be seen by asymptotic expansions for large values of  $|\alpha|$. For large $|\alpha|$
it is straightforward to check there is a solution of (\ref{zadef}) with  asymptotic expansion 
\begin{eqnarray*}
z_{\alpha,x} &\sim& 2\alpha^{1/2} - 2u_x\alpha^{-1/2}
- \alpha^{-3/2} \left( \frac12 u_{xxx} + u_x^2 \right) 
- \alpha^{-5/2} \left( \frac18 u_{xxxxx} + \frac32 u_xu_{xxx} + \frac54 u_{xx}^2 + u_x^3 \right) \\
&&  - \alpha^{-7/2} \left( \frac1{32} u_{xxxxxxx} +  \frac58 u_xu_{xxxxx} + \frac74 u_{xx}u_{xxxx} + \frac{19}{16}u_{xxx}^2 + \frac{15}{4}u_x^2u_{xxx} + \frac{25}{4} u_xu_{xx}^2
    + \frac54 u_x^4 \right)\\
&&  +\ldots\ . 
\end{eqnarray*}
This gives 
\begin{equation} 
Q(\alpha) =  \frac1{z_{\alpha,x}}   \sim  \frac12  \sum_{n=0}^\infty q_n \alpha^{-n-1/2} \ ,   \label{Qexp}
\end{equation}
where 
\begin{eqnarray}
  q_0   &=& 1\ ,\nonumber \\
  q_1   &=&  u_x \ ,  \nonumber\\
  q_2   &=& \frac14 u_{xxx} + \frac32 u_x^2 \ ,  \label{qs}  \\
  q_3   &=&  \frac1{16} u_{xxxxx} +   \frac54 u_xu_{xxx} + \frac58 u_{xx}^2 + \frac52 u_x^3 \ , \nonumber \\  
  q_4   &=&   \frac1{64} u_{xxxxxxx} + \frac7{16} u_xu_{xxxxx} + \frac78 u_{xx}u_{xxxx}  + \frac{21}{32} u_{xxx}^2 + \frac{35}{8} u_x^2u_{xxx} +
     \frac{35}{8} u_xu_{xx}^2  + \frac{35}{8} u_x^4   \ .  \nonumber \\
  &\vdots&     \nonumber 
\end{eqnarray}
Here we see explicitly the first few flows in the pKdV hierarchy.

Integrating with respect to $x$ and using (\ref{zat}) to fix the $t$-dependence of the constant of integration we obtain 
\begin{eqnarray}
  z_{\alpha} &\sim& 2\alpha^{3/2}t +   2\alpha^{1/2}x -  2u\alpha^{-1/2} 
  - {\alpha^{-3/2}} \left(I_1 +  \frac12 u_{xx} \right)
  - {\alpha^{-5/2}} \left(I_2 +  \frac18u_{xxxx} + \frac32 u_xu_{xx}   \right) \nonumber \\
 &&   - {\alpha^{-7/2}} \left(I_3 +  \frac1{32}u_{xxxxxx} + \frac58 u_xu_{xxxx} + \frac98 u_{xx}u_{xxx} + \frac{15}{4} u_x^2u_{xx}  \right)
  + \ldots    \ ,  \label{zexp} 
\end{eqnarray}
where
\begin{eqnarray*}
  I_1  &=& \int u_x^2 \ dx\ , \\   
  I_2  &=& \int \left( u_x^3-\frac14 u_{xx}^2  \right)\ dx \ , \\
  I_3  &=& \int \left( \frac54 u_x^4 - \frac54 u_xu_{xx}^2 + \frac1{16} u_{xxx}^2  \right) \ dx\ . 
\end{eqnarray*}
There remains the freedom to add  a constant of integration independent of $x$ and $t$, but this just corresponds to adding
a multiple of $Q(\alpha)$ to $R(\alpha)$, so we ignore it. Thus we obtain 
\begin{equation}
  R(\alpha) =  z_{\alpha,\alpha} Q(\alpha) - \frac32 t   \sim   \frac12  \sum_{n=0}^\infty r_n \alpha^{-n-1} \ ,    \label{Rexp}
\end{equation}
where 
\begin{eqnarray}
  r_0   &=& x q_0  + 3t q_1  \ , \nonumber \\   
  r_1   &=& x q_1  + 3t q_2 + u  \ ,  \nonumber \\
  r_2   &=& x q_2  + 3t q_3 + \frac32 I_1 q_0 + uu_{x} + \frac34 u_{xx}  \ , \label{rs} \\
  r_3   &=& x q_3  + 3t q_4 + \frac52 I_2 q_0 + \frac32 I_1 q_1 + \frac{9}{2}u_xu_{xx}+{\frac {5}{16}}u_{xxxx}+\frac{1}{4}uu_{xxx}+\frac{3}{2}uu_x^{2}\ , \nonumber \\
  r_4   &=& x q_4  + 3t q_5 + \frac72 I_3 q_0 + \frac52 I_2 q_1 + \frac32 I_1 q_2 + 18{u_x}^{2} u_{xx}+{\frac {33u_{xx}u_{xxx}}{8}} \nonumber \\
        &&  +  \frac{5}{2}u_xu_{xxxx}+{\frac {7}{64}}u_{xxxxxx}+\frac{1}{16}uu_{xxxxx}+\frac{5}{4}uu_xu_{xxx}+\frac{5}{8}uu_{xx}^{2}+\frac{5}{2}u{u_x}^{3}\ . \nonumber \\ 
  &\vdots&       \nonumber
\end{eqnarray}
Here we see explicitly the characteristics of Galilean symmetry, scaling symmetry and the first few
additional symmetries \cite{IS,OS,Khor}.   For general $n$, the terms that depend explicitly on $x$ and $t$
are $x q_n + 3 t q_{n+1}$, and a new integral term appears in each  $r_n$  for $n\ge 2$ (introducing
dependence on a new arbitrary constant). 

The components of $Q(\alpha)$ and $R(\alpha)$ can also be obtained by  use of the recursion operator.
This follows from the following result: 

\medskip\noindent
    {\bf Theorem 4}.
$Q(\alpha),S(\alpha),T(\alpha)$ are all solutions of the differential equation
$$  \left(  \textstyle{\frac14} \partial_x^3 + (2u_x-\alpha) \partial_x + u_{xx} \right) Q = 0 \ . $$      
$R(\alpha)$ is a solution of the differential equation
$$  \left(  \textstyle{\frac14} \partial_x^3 + (2u_x-\alpha) \partial_x + u_{xx} \right) R = -\textstyle{\frac12} \left( 1 + 3 tu_{xx} \right)\ .  $$      

\medskip
\noindent{\bf Proof:} A direct computation.   $\bullet$ 

\medskip
\noindent Substituting  the expansion (\ref{Qexp}) in the first result of the theorem
and equating coefficients we obtain 
$$  \left(  \textstyle{\frac14} \partial_x^3 + 2u_x \partial_x + u_{xx} \right) q_n =  q_{n+1,x}  \ , \qquad  n=0,1,\ldots\ , $$
and  $ q_0 = 1$.   Alternatively, this can be written
$$  q_{n+1} = \left(  \textstyle{\frac14} \partial_x^2 + \partial_x^{-1} u_x \partial_x + u_{x} \right) q_n   \ , \qquad  n=0,1,\ldots\ ,  $$
where on the right we see the recursion operator for symmetries of KdV \cite{Ol1,lenard}.  The recursion relation makes it easy
to find the  $Q_n$ in practice, but it is necessary to prove  that 
  the $Q_n$ constructed in this way are local (see, for example, \cite[Theorem 5.31]{Ol0}). Similarly, substituting the 
expansion (\ref{Rexp}) in the second result of the theorem we obtain
$$  \left(  \textstyle{\frac14} \partial_x^3 + 2u_x \partial_x + u_{xx} \right) r_n =  r_{n+1,x}  \ , \qquad  n=0,1,\ldots\ ,  $$
and $r_0 = x + 3tu_x$.  

 The  corresponding expansions of $S(\alpha)$ and $T(\alpha)$  for large $|\alpha|$ are  
\begin{eqnarray*}
  S(\alpha) &=&  \frac{e^{z_\alpha}}{z_{\alpha,x}}  \sim  \frac{e^{2 \alpha^{3/2} t + 2 \alpha^{1/2 } x}}{\alpha^{1/2}}\left( 1 + \sum_{n=1}^\infty s_n\alpha^{-n} \right)\ ,   \\
  T(\alpha) &=& \frac{e^{-z_\alpha}}{z_{\alpha,x}} \sim  \frac{e^{-2 \alpha^{3/2} t - 2 \alpha^{1/2 } x}}{\alpha^{1/2}}\left( 1 + \sum_{n=1}^\infty \tau_n\alpha^{-n} \right)   \ , 
\end{eqnarray*}
  where the $s_n$ and $\tau_n$ depend on $u$ and its $x$-derivatives and the integrals $I_1,I_2,\ldots$. However, as these expansions are not simple power
  series in $\alpha$, it is not clear if there is any sense in which they can be decomposed into an infinite hierarchy of ($\alpha$-independent) symmetries.

\subsection{Expansions for small values of $|\alpha|$} 

For small $\alpha$ we can find a formal power series solution of (\ref{zadef})-(\ref{zat}), i.e. a solution in the
form
$$ z_\alpha  = \sum_{n=0}^\infty  z_{n}(x,t)  \alpha^n \ ,   $$ 
giving power series expansions of $Q,R,S,T$ which we write, respectively, as
$$ \sum_{n=0}^\infty  Q_n(x,t)\alpha^n \ , \sum_{n=0}^\infty  R_n(x,t)\alpha^n \ , \sum_{n=0}^\infty  S_n(x,t)\alpha^n \ , \sum_{n=0}^\infty  T_n(x,t)\alpha^n\ .  $$
From theorem 4 we have 
$$  \left(  \textstyle{\frac14} \partial_x^3 + 2u_x \partial_x + u_{xx} \right) Q_{n+1}  =  Q_{n,x}  \ , \qquad  n=0,1,\ldots\ ,   $$
and 
$$  \left(  \textstyle{\frac14} \partial_x^3 + 2u_x \partial_x + u_{xx} \right) Q_{0}  =  0 \ .   $$    
The coefficients $S_n,T_n$ satisfy similar equations. Thus the associated hierarchies of symmetries can be constructed by application
of the {\em inverse} of the recursion operator, c.f. \cite{Guthrie,GH,Lou}.  When constructed in this way, at each step 3 arbitrary constants are introduced.
However, all the coefficients are fully determined  in terms of the  $z_n$. In particular we have
$$  Q_0 = \frac{1}{z_{0,x}} \ , \qquad S_0 = \frac{e^{z_0}}{z_{0,x}} \ , \qquad  T_0 = \frac{e^{-z_0}}{z_{0,x}}\ ,  $$
where $z_0$ is the solution of  (\ref{zadef})-(\ref{zat}) with $\alpha=0$. 

The coefficients $R_n$ satisfy 
$$  \left(  \textstyle{\frac14} \partial_x^3 + 2u_x \partial_x + u_{xx} \right) R_{n+1}  =  R_{n,x}  \ , \qquad  n=0,1,\ldots\ ,  $$
and are also constructed by applications of the inverse recursion operator. But the initial condition is given by   
$$  \left(  \textstyle{\frac14} \partial_x^3 + 2u_x \partial_x + u_{xx} \right) R_{0}  =  -\frac12 (1 + 3 t u_{xx})   \ . $$    
This has a particular solution
$$ R_0 = -\frac32 t  + \frac{2}{z_{0,x}} \int \frac{dx}{z_{0,x}} -  \frac{e^{-z_0}}{z_{0,x}} \int \frac{e^{z_0} dx}{z_{0,x}} -  \frac{e^{z_0}}{z_{0,x}} \int \frac{e^{-z_0} dx}{z_{0,x}}   $$ 
which can be found by application of the ``variation of constants'' method to the equation for $R_0$, using the fact that
$Q_0,S_0,T_0$ are linearly independent solutions of the associated homogeneous equation.

\subsection{Laurent expansions}  

In greater generality, results for the dependence of solutions of differential equations on a parameter guarantee that for suitable initial conditions, and
on any compact set, the solutions of  (\ref{zadef})-(\ref{zat}) will be analytic in $\alpha$ in a sufficiently small domain in the complex $\alpha$ plane. But 
typically there will be singularities. This is well illustrated by the case $u=0$, for which the general solution of the system is 
$$ z_\alpha = \log  \left(  \frac{C_1 e^{ \sqrt{\alpha} (x + \alpha t) } + C_2 e^{ -\sqrt{\alpha} (x + \alpha t) } }
                                {C_3 e^{ \sqrt{\alpha} (x + \alpha t) } + C_4 e^{ -\sqrt{\alpha} (x + \alpha t) } }
\right)\ ,  $$
where $C_1,C_2,C_3,C_4$ are constants (i.e. independent of $x$ and $t$, but not necessarily $\alpha$). Thus it is also natural to 
consider Laurent expansions, which we shall do in the next section. However,  for such expansions there are no explicit formulae for
the coefficients of the expansions of $Q,R,S,T$ in terms of the coefficients of the expansion of $z_\alpha$. 

\subsection{Commutators in components} 

We open this section with  explicit computations of the first few additional symmetries, as given in (\ref{rs}),  with
the standard symmetries, as given in (\ref{qs}), to emphasize the ambiguity in the commutators arising because of the nonlocal terms
in (\ref{rs}), or more specifically due to the need to define the action of the standard symmetries on  
the arbitrary constants in the integrals in (\ref{rs}). Direct computations (for $m=0,1,2,3,4$) give 
\begin{eqnarray} 
[r_0,q_m] &=& \left\{ \begin{array}{ll}   0  &  m=0 \\
                 (2m-1)q_{m-1}  &  m= 1,2,3,4  \end{array}\right. \ ,  \nonumber \\ 
{}[r_1,q_m]  &=&  (2m-1)q_{m}\ , \nonumber \\   
{}[r_2,q_m]  &=&  (2m-1)q_{m+1} + C_1 q_0\ ,  \label{rqcomms}\\  
{}[r_3,q_m]  &=&  (2m-1)q_{m+2} + C_1 q_1 + C_2 q_0\ ,  \nonumber  \\  
{}[r_4,q_m]  &=&  (2m-1)q_{m+3} + C_1 q_2 + C_2 q_1 + C_3 q_0\ ,   \nonumber
\end{eqnarray}
where $C_1,C_2,C_3$ are free constants. Taking these to be zero, these formulae are consistent with the traditional result \cite{MZ,OFB,ASvM} 
$$ [r_n,q_m] = \left\{
\begin{array}{cc}
  (2m-1) q_{m+n-1}  & n+m > 0 \\ 
  0                & n=m=0 
\end{array}
\right.\ .  $$
The traditional result for the commutators of the additional symmetries \cite{OFB,MZ,ASvM}  is  
$$  [r_n, r_m ] =  2(m-n) r_{m+n-1}\ .    $$
This corresponds to a specific choice of a different set of arbitrary constants, associated with the choice of the action
of the $r_n$ on the integrals $I_m$ appearing in (\ref{rs}). We now show that these
commutators  can be derived  from  the third of the commutator algebras for $Q,R$ given in section 3.  

\medskip 
\noindent{\bf Theorem 5:} Assume that  $Q,R$ have asymptotic expansions
\begin{equation}
Q(\alpha) \sim \frac12 \sum_{n=0}^\infty  q_n\alpha^{-n-1/2}  \ , \qquad
R(\alpha) \sim \frac12 \sum_{n=0}^\infty  r_n\alpha^{-n-1}  \ .   \
\label{exps}\end{equation}
Then the commutator algebra of theorem 3$''$ implies
\begin{eqnarray}
&&  [q_n,q_m] =   0 \ , \label{qqcomm}\\ 
&&  [r_n,q_m] = \left\{ \begin{array}{cc}
  (2m-1) q_{n+m-1}  & n+m > 0 \\ 
  0                & n=m=0 
\end{array}    \right.    \ ,  \label{rqcomm} \\    
&&  [r_n,r_m] =  2(m-n) r_{m+n-1}   \ ,    \label{rrcomm} 
\end{eqnarray}
for $n,m\ge 0$.   

\medskip 
\noindent{\bf Notation:} We write
$$ H_n =  \left\{ \begin{array}{ll} 1 & n \ge 0 \\  0  & n < 0 \end{array} \right.  \ .   $$ 
We write
$$ \sigma_{nm} =  \left\{ \begin{array}{ll} 1 & n,m \ge 0 \\
                                       -1 & n,m < 0 \\
                                      0 & {\rm otherwise}  \end{array} \right. \ .  $$
We have
$$  \sigma_{nm} =  H_n - H_{-m-1} \ .   $$

\medskip
\noindent{\bf Proof:} We present the calculations assuming that $|\beta|>|\alpha|$. Similar calculations assuming $|\beta|<|\alpha|$ lead to
the same result. Using the $R,Q$ commutator (\ref{RQ3}) we have
\begin{eqnarray*}
&& \frac14  \sum_{n=0}^\infty \sum_{m=0}^\infty [r_n,q_m] \alpha^{-m-1/2}  \beta^{-1-n}  \\ 
  &=&  [R(\beta),Q(\alpha)]   \\
  &=&   \frac{\partial}{\partial\alpha}\left( \frac{Q(\alpha)}{\alpha-\beta} \right) +  \frac{(3\beta-\alpha)\alpha^{1/2}}{2\beta^{3/2}(\beta-\alpha)^2} Q(\beta)    \\
  &=&   \frac{\partial}{\partial\alpha}\left(
  -\frac1{2\beta} \left( \sum_{M=0}^\infty q_M \alpha^{-M-1/2}  \right) \left( \sum_{N=0}^\infty \left( \frac{\alpha}{\beta} \right)^{N} \right)  
  \right)  \\
  &&  +   \frac{(3\beta-\alpha)\alpha^{1/2}}{4\beta^{7/2}}    
   \left( \sum_{M=0}^\infty q_M \beta^{-M-1/2}  \right) \left( \sum_{N=0}^\infty N \left( \frac{\alpha}{\beta} \right)^{N-1} \right)   \\
   &=&   -\frac12  \sum_{M=0}^\infty \sum_{N=0}^\infty   \left( N-M-\frac12  \right) q_M \alpha^{N-M-3/2} \beta^{-N-1}  \\
   &&  + \frac34  \sum_{M=0}^\infty \sum_{N=0}^\infty   N q_M \alpha^{N-1/2} \beta^{-M-N-2}
   - \frac14  \sum_{M=0}^\infty \sum_{N=0}^\infty   N q_M \alpha^{N+1/2} \beta^{-M-N-3}     \\
   &=&  \frac14  \sum_{n=-\infty}^\infty \sum_{m=-\infty}^\infty    q_{m+n-1} \alpha^{-m-1/2} \beta^{-1-n}  H_{n+m-1}
   \left(   (2m-1) H_n    -  3 m H_{-m}  + (m+1) H_{-m-1}     \right)   \ . 
\end{eqnarray*}  
In moving from the penultimate line to the last line we have substituted
$M=n+m-1,N=n$ in the first series, 
$M=n+m-1,N=-m$ in the second, and
$M=n+m-1,N=-m-1$ in the third.  Since  $m H_{-m} = m H_{-m-1} $, the last term can be simplified, and the right hand side takes the
form 
$$   \frac14  \sum_{n=-\infty}^\infty \sum_{m=-\infty}^\infty   (2m-1)  q_{m+n-1} \alpha^{-m-1/2} \beta^{-1-n}  H_{n+m-1}  \left(  H_n    -  H_{-m-1}  \right)   \ .  $$
  Finally, since $  H_n    -  H_{-m-1}  = \sigma_{nm}$,  the right hand side can be written 
$$   \frac14  \sum_{n=0}^\infty \sum_{m=0}^\infty   (2m-1)  q_{m+n-1} \alpha^{-m-1/2} \beta^{-1-n}  H_{n+m-1}   \ .  $$
  Comparing the left and right hand sides we obtain (\ref{rrcomm})   as required.
The calculation for the $[r_n,r_m]$ commutator is very similar, using the $R,R$ commutator (\ref{RR3}), and we do not present it.
(A slightly more general calculation is presented in full in the proof of the next theorem.)
$\bullet$   
  
\medskip\noindent{\bf Notes:}
\newline 1. It is clear that it is very fortuitous that it is possible to substitute the expansions (\ref{exps}) into the
commutator (\ref{RQ3}) and obtain a consistent commutator algebra for the components. A slight change in the form of 
(\ref{RQ3}) could make the calculation presented above inconsistent, and it is also not at all clear, {\it ab initio}, that
the same results will be obtained in both the cases $|\beta|>|\alpha|$ and $|\beta|<|\alpha|$. With regard to the latter, we
suspect that this is related to the fact that the variations of $z_\alpha$ under the symmetries $Q(\beta)$ and $R(\beta)$ used 
in theorem 3$''$, equations (\ref{deltaQ3})-(\ref{deltaR3}), can be continued also for the case $\beta=\alpha$, as opposed to
some of the variations used in theorem 3, specifically equation (\ref{deltaQ}). For the commutator algebra of theorem 3 we do not know
any way to expand $Q$ and $R$ and obtain a consistent algebra for the components.  The commutator algebra of theorem 3$'$, however,
is consistent with Laurent expansions of $Q,R,S,T$, as we shall show below.
  \newline 2. In the algebra (\ref{qqcomm})-(\ref{rrcomm}), the nonlocal symmetry $r_2$ is a {\em master symmetry} \cite{fuchs},\cite[Section 5.2]{Ol0}--- starting
  from $q_0$ and repeatedly computing commutators with $r_2$ generates all the symmetries $q_n$. 

\medskip 
\noindent{\bf Notation:}   In the following theorem  we use  $Q_n,R_n,S_n,T_n$ as components of  $Q,R,S,T$ in Laurent expansions.
In section 4.2 we used the same notation for the components in Taylor expansions, which is a special case. 

\medskip 
\noindent{\bf Theorem 6:} Assume that in a given annulus $Q,R,S,T$ have Laurent expansions
$$ \sum_{n=-\infty}^\infty  Q_n\alpha^n \ , \sum_{n=-\infty}^\infty  R_n\alpha^n \ , \sum_{n=-\infty}^\infty  S_n\alpha^n \ , \sum_{n=-\infty}^\infty  T_n\alpha^n\ .  $$
Then the commutator algebra of theorem 3$'$ implies
\begin{eqnarray}
&&  [Q_n,Q_m] =  [S_n,S_m] =   [T_n,T_m]  =   0 \ , \\ 
&&  [Q_n,S_m] =  S_{m+n+1} \sigma_{nm}     \ ,  \\    
&&  [Q_n,T_m] =  -T_{m+n+1} \sigma_{nm}     \ ,  \\
&&  [T_n,S_m] =  2Q_{m+n+1} \sigma_{nm}     \ ,  \\    
&&  [R_n,Q_m] =  (m+1)Q_{m+n+2} \sigma_{nm}     \ ,  \label{RQcomm}\\    
&&  [R_n,S_m] =  (m+1)S_{m+n+2} \sigma_{nm}     \ ,  \\        
&&  [R_n,T_m] =  (m+1)T_{m+n+2} \sigma_{nm}     \ ,  \\        
&&  [R_n,R_m] =  (m-n) R_{m+n+2} \sigma_{nm}     \ ,  
\end{eqnarray}
for $n,m\in {\bf Z}$.   

\medskip
\noindent{\bf Proof:} In all the following calculations we assume  $|\beta|<|\alpha|$. The same results are obtained if we
assume  $|\beta|>|\alpha|$. 
\newline 1. From the $[Q(\beta),S(\alpha)]$ commutator (\ref{QS2}):
\begin{eqnarray*}
&&  \sum_{m=-\infty}^\infty \sum_{n=-\infty}^\infty [Q_n,S_m] \alpha^m  \beta^n  \\ 
  &=&  [Q(\beta),S(\alpha)]   \\
  &=&  \frac{ S(\alpha)- S(\beta)  }{\alpha-\beta}   \\ 
  &=&  \frac1{\alpha}  \left( \sum_{M=-\infty}^\infty  S_M (\alpha^M-\beta^M) \right)  \sum_{N=0}^\infty \left(\frac{\beta}{\alpha}\right)^N   \\
&=&  \sum_{M=-\infty}^\infty \sum_{N=0}^\infty S_M     \alpha^{M-N-1}\beta^N
   -  \sum_{M=-\infty}^\infty \sum_{N=0}^\infty S_M     \alpha^{-N-1}\beta^{M+N}   \\
&=&  \sum_{m=-\infty}^\infty \sum_{n=-\infty}^\infty S_{m+n+1} H_n     \alpha^m\beta^n 
   -  \sum_{m=-\infty}^\infty \sum_{n=-\infty}^\infty S_{m+n+1} H_{-m-1}      \alpha^m\beta^n \ ,   \\ 
\end{eqnarray*}  
where in the first sum we have substituted  $N=n$,  $M=m+n+1$ and in the second  sum we have substituted  $N=-m-1$,  $M=m+n+1$.
Comparing coefficients of  $\alpha^m \beta^n$ gives
$$ [Q_n,S_m] =  S_{m+n+1} (H_n - H_{-m-1} )  =  S_{m+n+1}\sigma_{nm}\ .   $$ 
Calculations for the $[Q,T]$ and $[T,S]$ commutators are similar. 
\newline 2. From the $[R(\beta),Q(\alpha)]$ commutator (\ref{RQ2}):
\begin{eqnarray*}
&&  \sum_{m=-\infty}^\infty \sum_{n=-\infty}^\infty [R_n,Q_m] \alpha^m  \beta^n  \\ 
  &=&  [R(\beta),Q(\alpha)]   \\
  &=&   \frac{\partial}{\partial\alpha}\left( \frac{Q(\alpha)}{\alpha-\beta} \right) +  \frac{Q(\beta)}{(\alpha-\beta)^2}   \\
  &=&   \frac{\partial}{\partial\alpha}\left(
  \frac1{\alpha} \left( \sum_{M=-\infty}^\infty Q_M \alpha^M  \right) \left( \sum_{N=0}^\infty \left( \frac{\beta}{\alpha} \right)^{N} \right)  
  \right)
  +  \frac1{\alpha^2} \left( \sum_{M=-\infty}^\infty Q_M \beta^M  \right) \left( \sum_{N=0}^\infty N \left( \frac{\beta}{\alpha} \right)^{N-1} \right)   \\
  &=&  \sum_{M=-\infty}^\infty \sum_{N=0}^\infty (M-N-1)  Q_M \alpha^{M-N-2}  \beta^N 
  +  \sum_{M=-\infty}^\infty \sum_{N=0}^\infty N Q_M \alpha^{-N-1}\beta^{M+N-1}     \\
  &=&   \sum_{m=-\infty}^\infty \sum_{n=-\infty}^\infty (m+1)  Q_{m+n+2} H_n     \alpha^m\beta^n
  +   \sum_{m=-\infty}^\infty \sum_{n=-\infty}^\infty  (-m-1) Q_{m+n+2} H_{-m-1}  \alpha^m\beta^n \ , 
\end{eqnarray*}  
where in the first sum we have substituted  $N=n$,  $M=m+n+2$ and in the second  sum we have substituted  $N=-m-1$,  $M=m+n+2$.  
Comparing coefficients of  $\alpha^m \beta^n$ gives
$$ [R_n,Q_m] =  (m+1)Q_{m+n+2} (H_n - H_{-m-1} )  =  (m+1) Q_{m+n+2}\sigma_{nm}\ .   $$ 
Calculations for the $[R,S]$ and $[R,T]$ commutators are similar. 
\newline 3. From the $[R(\beta),R(\alpha)]$ commutator (\ref{RR}): 
\begin{eqnarray*}
&&  \sum_{m=-\infty}^\infty \sum_{n=-\infty}^\infty [R_n,R_m] \alpha^m  \beta^n  \\ 
&=&  [R(\beta),R(\alpha)]   \\
&=&  \frac{2(R(\beta)-R(\alpha))}{(\alpha-\beta)^2}  + \frac{R'(\beta)+R'(\alpha)}{\alpha-\beta} \\
&=& \frac{2}{\alpha^2} \left(  \sum_{N=-\infty}^\infty   R_N (\beta^N - \alpha^N) \right)  \left(  \sum_{M=0}^\infty  M \left( \frac{\beta}{\alpha} \right)^{M-1} \right)    \\
&& +  \frac{1}{\alpha}  \left(  \sum_{N=-\infty}^\infty   NR_N (\beta^{N-1} + \alpha^{N-1}) \right)  \left(  \sum_{M=0}^\infty  \left( \frac{\beta}{\alpha} \right)^{M} \right)    \\
&=&   \sum_{N=-\infty}^\infty \sum_{M=0}^\infty 2 M R_N  \alpha^{-M-1} \beta^{M+N-1} -   \sum_{N=-\infty}^\infty \sum_{M=0}^\infty 2 M R_N  \alpha^{N-M-1} \beta^{M-1} \\
  &&  + \sum_{N=-\infty}^\infty \sum_{M=0}^\infty N R_N  \alpha^{-M-1} \beta^{M+N-1}  +   \sum_{N=-\infty}^\infty \sum_{M=0}^\infty N R_N  \alpha^{N-M-2} \beta^{M}  \\
  &=&   \sum_{n=-\infty}^\infty  \sum_{n=-\infty}^\infty 2(-1-m) H_{-m-1 } R_{n+m+2}  \alpha^m \beta^n 
  -  \sum_{n=-\infty}^\infty  \sum_{n=-\infty}^\infty  2 (n+1) H_{n+1} R_{n+m+2}   \alpha^m \beta^n  \\
&&    +  \sum_{n=-\infty}^\infty  \sum_{n=-\infty}^\infty  (m+n+2) H_{-m-1} R_{n+m+2} \alpha^m \beta^n 
  + \sum_{n=-\infty}^\infty  \sum_{n=-\infty}^\infty   (m+n+2) H_{n} R_{n+m+2} \alpha^m \beta^n 
\end{eqnarray*}
Since $(n+1)H_{n+1} = (n+1) H_n  $, by comparing coefficients we have
$$ [R_n,R_m] =  (m-n) ( H_{n}  - H_{-m-1} )  R_{n+m+2}=  (m-n)  R_{n+m+2} \sigma_{nm}\ .  \bullet  $$ 

\medskip\noindent{\bf Notes:}
\newline 1. Remarkably, due to the appearance of the  $\sigma_{nm}$ term in all the nontrivial commutators, the
algebra of symmetries described in this theorem decomposes into the direct sum of two algebras associated with the
negative and the non-negative modes.
\newline 2. Our results should be compared with those of Guthrie \cite{Guthrie} and Lou \cite{Lou}. While there is a certain commonality,  
there are major differences. Guthrie claims there is a $sl(2,{\bf R})$ loop algebra of symmetries. Due to the $\sigma_{nm}$
terms in the commutators, the algebra we have identified
associated with the $Q,S,T$ symmetries is not a full loop algebra, 
but the direct sum of two copies of the analytic loop algebra (i.e
the algebra of maps of the unit circle into $sl(2,{\bf R})$ that are the boundary values of analytic functions on the unit
disk).
\newline 3. It should be emphasized that none of the symmetries $Q_n,S_n,T_n$ described in this section can be
associated with the standard symmetries $q_n$ of the KdV equation, which  appear as coefficients of the asymptotic series 
(\ref{Qexp}) which is not directly related to the Laurent series for $Q$ used in this section. However, we can associate the
additional symmetries $r_n$ with the negative modes in the Laurent expansion for $R$.  Specifically we have
$$   r_n =   2R_{-1-n} \ , \qquad  n=0,1,\ldots \ .   $$
With this identification the last commutator in theorem 6 gives the commutator (\ref{rrcomm}). Note, however, the
distinction between the $[R_n,Q_m]$ commutator (\ref{RQcomm}) and the $[r_n,q_m]$ commutator (\ref{rqcomm}).
But $R_{-3}$ still plays the role of a master symmetry. Taking repeated commutators of $Q_{-2},S_{-2},T_{-2}$
with $R_{-3}$ generates all the symmetries $Q_n,S_n,T_n$ for  $n<-2$.

\section{Relation to B\"acklund transformations} 

In this section we show that the symmetries  $Q(\alpha),S(\alpha),T(\alpha)$ are all infinitesimal double B\"acklund transformations
(BTs). We recall the following argument from  \cite{RSG}.  
The pKdV equation (\ref{pKdV}) has the B\"acklund transformation $u\rightarrow u+v_\alpha$ where  $v_\alpha$ satisfies
\begin{eqnarray}
v_{\alpha,x} &=&\alpha-2u_x -v_\alpha^2 \ ,  \label{BT10}\\
v_{\alpha,t} &=&-\textstyle{\frac12} u_{xxx}  + (\alpha+u_x)(\alpha -2u_x-v_\alpha^2) + u_{xx}v_\alpha \ ,  \label{BT11}
\end{eqnarray}
and $\alpha$ is a parameter.  BTs are known to commute. If $u_\alpha = u + v_\alpha$ is the solution found by applying a BT with parameter $\alpha$
and  $u_\beta = u + v_\beta$ is the solution found by applying a BT with parameter $\beta$, then the solutions found by first applying
the BT with parameter $\alpha$ and then the BT with parameter $\beta$, or by applying the two transformations in reverse order, are
the same, and are given by the ``superposition principle''
$$  u_{\alpha\beta} = u_{\beta\alpha}  =  u + \frac{\beta-\alpha}{u_\beta-u_\alpha} \ .   $$ 
Note that applying a BT to a given solution of KdV, for a given value of the parameter, does not generate a single solution, but rather a
1-parameter family, because of the constant of integration that appears when solving (\ref{BT10})-(\ref{BT11}). Thus we can consider the
superposition principle in the limit  $\beta \rightarrow \alpha$ without $u_\beta \rightarrow u_\alpha$.
 Writing $u_\alpha^{(1)}$  instead of $u_\alpha$ and taking $\beta = \alpha + \epsilon$,
  we have $u_\beta = u_\alpha^{(2)} + O(\epsilon)$, where $u_\alpha^{(2)} \not= u_\alpha^{(1)}$,   and thus
  $$  u_{\alpha\beta} = u_{\beta\alpha}  =  u + \frac{\epsilon}{u_\alpha^{(2)}-u_\alpha^{(1)}} + O(\epsilon^2)\ .   $$ 
Thus  in this limit we obtain an infinitesimal symmetry of pKdV with characteristic
$$  \frac{1}{ u^{(1)}_\alpha - u^{(2)}_\alpha} \ ,$$ 
where  $u^{(1)}_\alpha,u^{(2)}_\alpha$ are two distinct solutions of pKdV obtained by applying the BT with parameter $\alpha$ to the solution $u$. 
It is straightforward to check that if  $z_\alpha$ satisfies  the system  (\ref{zadef})-(\ref{zat}) then
\begin{equation}
v^{(1)}_{\alpha} = \frac12 \left( z_{\alpha,x} -   \frac{z_{\alpha,xx}}{z_{\alpha,x}}  \right) \ , \qquad
     v^{(2)}_{\alpha} = \frac12 \left( - z_{\alpha,x} -  \frac{z_{\alpha,xx}}{z_{\alpha,x}}  \right)   \label{vtoz}   
\end{equation}
are two solutions of (\ref{BT10})-(\ref{BT11}). In this way we see the characteristic of the symmetry associated with an infinitesimal
double BT as just described  is simply
$$Q(\alpha) = \frac{1}{ v^{(1)}_\alpha - v^{(2)}_\alpha} = \frac{1}{z_{\alpha,x}}\ . $$ 

The superposition principle describes the entire two parameter family of solutions obtained by applying two BTs with different parameters
$\alpha$ and $\beta$ (in either order). But in the case $\beta=\alpha$ it gives the faulty impression that only a single solution is
generated by a double BT, namely the original solution $u$. This is incorrect. A more careful analysis requires considering the limit
$\beta \rightarrow \alpha$ {\it with} $u_\beta \rightarrow u_\alpha$ and applying l'H\^opital's rule. This gives the following result which
can also be directly verified: 

\medskip 
\noindent{\bf Theorem 7:} If $u$ is a solution of pKdV and $p$ is a solution of the system 
\begin{eqnarray}
p_{xx} &=&(\alpha-2u_x)p   -   2p^2p_x - \textstyle{\frac14}  p^5  \ ,  \label{pxx}\\
p_{t} &=& {\textstyle{\frac12}} pp_x^2 +  \left(\alpha+u_x +  {\textstyle{\frac12}} p^4 \right) p_x   
   - {\textstyle{\frac12}} pu_{xx} - {\textstyle{\frac12}} p^3 ( \alpha - 2u_x)   + {\textstyle{\frac18}} p^7  
\ .  \label{pt}   
\end{eqnarray}
then $u+p^2$ is also a solution of pKdV.   

\medskip\noindent A form of this transformation appears in \cite{galas}.  
Transformations of this sort (which are double BTs with a fixed parameter $\alpha$) can be superposed 
with regular BTs and with themselves and the order in which they are applied is not important. We give just
one example of a superposition principle:  

\medskip 
\noindent{\bf Theorem 8:} 
If $u\rightarrow u+p^2$ is the double BT with parameter $\alpha$ and 
If $u\rightarrow u+q^2$ is the double BT with parameter $\beta$, then their superposition is
$$ u \rightarrow u + p^2 + q^2 - \frac{W( (p^2+q^2)W + 2pq(\alpha-\beta) )}{W^2 - (\alpha-\beta)^2} \ ,   $$ 
where
$$  W = pq_x - qp_x + \textstyle{\frac12} pq(q^2-p^2) \ .  $$

\medskip\noindent
It is also straightforward to check that if $u\rightarrow u+p^2$ is a double BT with parameter $\alpha$, then
$$ u \rightarrow u + \frac{p_x}{p} + \frac12 p^2 $$
is a  standard BT with parameter $\alpha$.  

We note that $p=0$ is a solution of the system (\ref{pxx})-(\ref{pt}), reflecting the fact that the two-parameter
family of solutions generated by a double BT always includes the original solution.  We can also consider the limit
of small $p$. In this way we get an infinitesimal symmetry  $u \rightarrow u+p^2$ where $p$ satisfies the linear
system 
\begin{eqnarray*}
p_{xx} &=&(\alpha-2u_x)p \ , \\
p_{t} &=&   \left(\alpha+u_x \right) p_x  - {\textstyle{\frac12}} pu_{xx} \ . 
\end{eqnarray*}
Comparing with note 3 after theorem 1 we conclude  that {\it the infinitesimal symmetries $S(\alpha)$ and $T(\alpha)$ are 
  infinitesimal versions of double BTs.}

This however leads to an apparent conundrum. We know that BTs commute, and have identified $Q(\alpha),S(\alpha),T(\alpha)$ 
as infinitesimal BTs. So surely this means that $Q,S,T$ should mutually commute? Indeed, in the algebra of theorem 3, they commute.
But in the algebra of theorem 3$'$ they do not. The resolution of this lies in the fact that BTs do not generate a single solution,
but a manifold of solutions. The statement that ``BTs commute'' means that the entire manifold of solutions obtained by applying
2 BTs is independent of the order in which they are applied. But if we select a single solution each time we apply a BT,
according to some rule, there is no guarantee we will end up with the same one when we apply BTs in a different order. This is in
analogy with the fact we have seen  that to define the commutator algebra of nonlocal symmetries it is necessary to define the action of the
symmetries on the integrals or prepotentials used to define them, and this can be done in different ways, giving distinct algebras. 

\section{Concluding remarks}

Quite possibly the most significant contribution of our results is to the theory of nonlocal symmetries. We have given an explicit 
example of the possibility of  distinct commutator algebras  being associated with a single set of nonlocal symmetries. This is
a possibility that seems to have been overlooked in the literature and merits much further investigation. The three versions
of the algebra that we have looked at all seem to be of some importance. In the first, the $Q,S,T$ symmetries commute, reflecting their
origin as infinitesimal double B\"acklund transformations. In the second, the coefficients of Laurent expansions of $Q,S,T$ satisfy
the commutation relations of an $sl(2,{\bf R})$ loop group, which is known to play a significant role in the theory of the KdV hierarchy,
through the Segal-Wilson correspondence \cite{SW}. However, the well-known commutators of the most prominent symmetries of KdV, the standard hierarchy
and the hierarchy of additional symmetries, emerge from the third version of the algebra. 

From the point of view of the KdV equation,  all known infinitesimal symmetries seem to be encapsulated in the
4 generating symmetries $Q,R,S,T$
 and this would seem to be a great simplification, despite the issues arising from nonlocality.
$Q,S,T$ were known before, albeit in a different form, but  as far as we know the expression  (\ref{Rdef}) for $R$ is new.
From the algebra in which $Q,S,T$ commute we deduced the possibility of adding further (nonlocal) flows to the KdV hierarchy; we are also unaware
of any previous suggestion of this possibility. It would be interesting 
to see whether the actions of $Q,R,S,T$ can be expressed succinctly in terms of the tau function.
We note that  in analogy to the definition of integrability for finite dimensional Hamiltonian systems, 
  the integrability of KdV is often attributed purely to the existence of an infinite hierarchy of
  commuting symmetries, the symmetries we have called $q_n$. It is clear there is a far richer symmetry structure than this.

  Finally, we reiterate a point made in section 2, that although we have written the generating symmetries of KdV in terms
  of the single field $z_\alpha$, they can also be written in terms of a pair of solutions of the Lax pair, and this gives an
  obvious way to search for generating symmetries for other $1+1$-dimensional integrable equations \cite{SashaPreprint}.  Thus we
  are hopeful that the program fully implemented here for the KdV equation can also be implemented in other cases.

\section*{Data Availability}
Data sharing is not applicable to this article as no datasets were generated or analysed during the current study.

\def\cprime{$'$} \def\cprime{$'$}

\end{document}